\documentclass{eas}
\usepackage{graphicx}
\begin{document}
\title{Dust-forming molecules in VY Canis Majoris (and Betelgeuse)} 
\author{T. Kami\'{n}ski}\address{Max-Planck Institut f\"ur Radioastronomie, 
       Auf dem H\"ugel 69, 53121 Bonn, Germany}
\author{C. A. Gottlieb}\address{Harvard-Smithsonian Center for Astrophysics, 60 Garden Street, Cambridge, MA, USA}
\author{M. R. Schmidt}\address{Nicolaus Copernicus Astronomical Center, Rabia\'nska 8, 87-100 Toru\'n, Poland}
\author{N.~A.~Patel$^2$}
\author{K.~H.~Young$^2$}
\author{K. M. Menten$^1$}
\author{S. Br{\"u}nken}\address{I. Physikalisches Institut, Z\"ulpicher Strasse 77, 50937 K\"oln, Germany}
\author{H.~S.~P.~M{\"u}ller$^4$}
\author{J.~M.~Winters}\address{IRAM, 300 rue de la Piscine, 38406 Saint-Martin d'H{\'e}res, France}
\author{M. C. McCarthy$^2$}
\runningtitle{Dust-forming molecules in VY\,CMa (and Betelgeuse)}
\begin{abstract}
The formation of inorganic dust in circumstellar environments of evolved stars is poorly understood. Spectra of molecules thought to be most important for the nucleation, i.e. AlO, TiO, and TiO$_2$, have been recently detected in the red supergiant VY\,CMa. These molecules are effectively formed in VY\,CMa and the observations suggest that non-equilibrium chemistry must be involved in their formation and nucleation into dust. In addition to exploring the recent observations of VY\,CMa, we briefly discuss the possibility of detecting these molecules in the ``dust-poor" circumstellar environment of Betelgeuse.      
\end{abstract}
\maketitle
\section{Introduction}
Evolved stars are primary producers of dust in galaxies. They are sources of both the carbon-based dust,  originating mainly from carbon-rich asymptotic giant branch stars, and inorganic dust,  formed in oxygen-rich circumstellar environments including those of red supergiants. Despite the important role of dust in a broad range of astrophysical phenomena, the process of dust formation in circumstellar environments is poorly understood. In general, this process is a chain of chemical reactions starting from small gas-phase molecules which form successively larger molecules; these grow, form clusters, and end as macroscopic complexes with solid-state properties. The inorganic dust formation is likely to start from oxides. Because the nucleation occurs at rather high temperatures ($\sim$1000--1200\,K), these oxides must be refractory. Also, they should have high nucleation rates and be abundant enough to be able to effectively form dust. After ruling out more abundant elements (Si, Fe, and Mg), it was concluded that oxides of titanium (TiO, TiO$_2$) and -- to a lesser degree -- those of aluminum (AlO) should be the most important gas-phase species that initiate the formation of clusters (the so-called {\em seeds}) in oxygen-rich environments (Gail \& Sedlmayr \cite{GS}). The widely known ``silicates", which determine the observed properties of warm and cold dust in circumstellar shells, are important for the grain growth at lower temperatures (a few hundred K).
   
The gas-phase oxides of titanium and aluminum are very rarely observed in {\em circumstellar} environments of oxygen-rich stars. The electronic bands of the two monoxides, TiO and AlO, are well known to be present in optical spectra of photospheres of late type stars, but features that arise from further away from the star, i.e. in its circumstellar envelope, have been seen so far only in spectra of some Miras (e.g., Keenan \etal\ \cite{miras_alo}; see also Smolders \etal\ \cite{sm12}) and peculiar explosive variables (young stellar objects, red novae and related objects) (e.g., Kami\'nski \etal\ \cite{kami_v4332}). Even when TiO and AlO are observed, the exact location and the physical state of the emitting gas have not been well constrained hampering the observational verification of the dust-nucleation scenario. The only exception where such verification can now be attempted is the red supergiant VY~Canis Majoris (VY\,CMa). 

Recent years brought discoveries of previously unobserved molecules in the millimeter spectrum of VY\,CMa (Tenenbaum \etal\ \cite{T10} and references therein). In the optical, the object has been long known for its extraordinary spectrum containing molecular bands of TiO, VO, ScO, and YO seen {\em in emission} (Hyland \etal\ \cite{hyland}; Wallerstein \cite{w86,w71}; Herbig \cite{herbig}). It is not an exaggeration to state that VY\,CMa has the richest molecular spectrum among all currently known oxygen-rich evolved stars. It has a high mass-loss rate ($\sim$10$^{-4}$\,M$_{\odot}$\,yr$^{-1}$), which produces a complex emission and reflection nebula (Humphreys \etal\ \cite{H05,H07}). It is therefore the primary target at all wavelengths for searches of gas-phase molecules that are important for the formation of inorganic dust. We have searched and successfully identified emission of TiO, TiO$_2$, and AlO in the optical and submillimeter spectra of VY\,CMa. Below we characterize the emission and briefly describe the possible implications of these findings.

\section{The dust-forming oxides in VY\,CMa}
\subsection{Titanium oxides in VY\,CMa}
\begin{figure*}\centering
\includegraphics[angle=270,width=\textwidth]{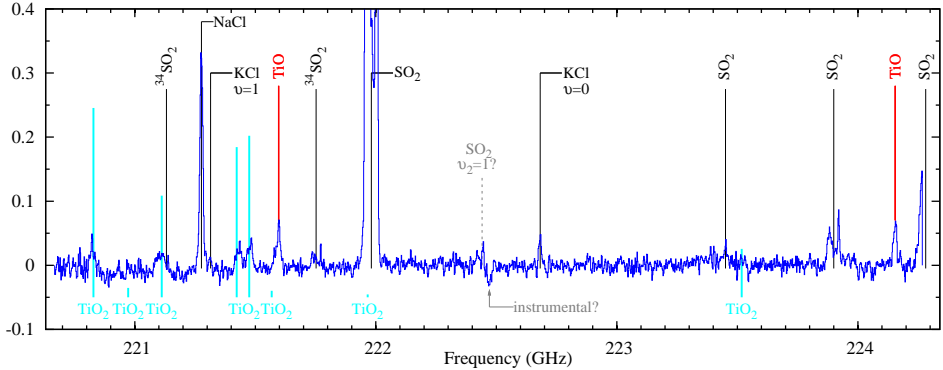}
\caption{The spectrum of VY\,CMa obtained with PdBI where features of TiO (red labels) and of TiO$_2$ (cyan) are clearly detected. The height of the lines marking TiO$_2$ positions, corresponds to the \emph{relative} intensities in LTE. The ordinate is flux density in Jy/beam.}
\label{Fig-PdBI}
\end{figure*}

In 2010, we used the Submillimeter Array (SMA) to obtain a line survey toward VY\,CMa in the range 279--355\,GHz (0.844--1.075\,mm). These were the most sensitive broad-bandwidth submillimeter observations of VY\,CMa to date. We detected $\sim$220 lines from 20 different molecules, some of which are the first detections of the species in this object or even in any astrophysical object. In particular, we found lines of TiO and TiO$_2$. Their presence was confirmed in 2012 with IRAM's Plateau de Bure Interferometer (PdBI). The spectrum obtained with PdBI is shown in Fig.\,\ref{Fig-PdBI}. These are the first detections of the two molecules in the ``radio" regime (TiO being detected before only in the optical), and, in case of TiO$_2$, it is the first (and only) astronomical observation of the molecule so far.

\begin{figure*}\centering
\includegraphics[angle=270,width=\textwidth]{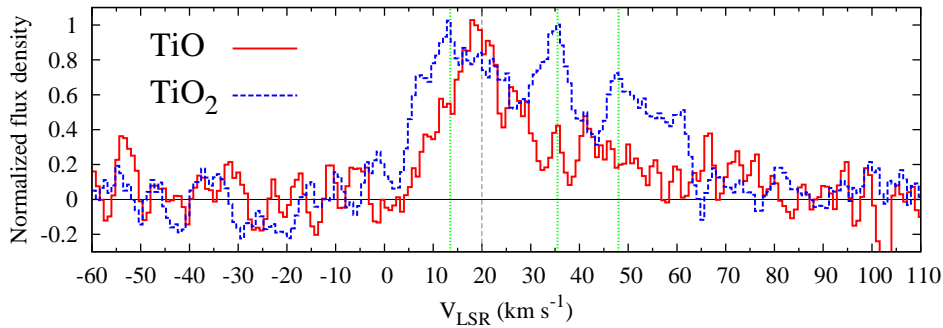}
\caption{The average profiles of TiO (red, solid line) and TiO$_2$ (blue, dashed line) observed with SMA. The vertical dashed line (grey) marks the stellar systemic velocity, while the vertical dotted (green) lines mark three stronest velocity components in the profile of TiO$_2$. These average profiles do not include lines observed with PdBI.}
\label{Fig-aver}
\end{figure*}

The lines of TiO are detected at a low signal-to-noise ratio (S/N), but the combined profile shows that the emission consists of a central narrow component with a half-width of FWHM/2\,=\,9\,km\,s$^{-1}$ and a broad pedestal of emission reaching the outflow terminal velocity of 40\,km\,s$^{-1}$ (Fig.\,\ref{Fig-aver}). Most of the emission therefore arises in the wind acceleration zone, which was observed earlier at very high spatial resolution in maser lines of H$_2$O and SiO. From an expansion model constrained by maser observations (Richards \etal\ \cite{anita}), we estimated that the TiO emission region extends to $r$\,$\approx$\,30\,R$_{\star}$ (R$_{\star}$=2000\,R$_{\odot}$) from the star. Some amount of TiO must extend above this region to give  rise to the broad pedestal. At our sensitivity, the TiO emission is not resolved, so it is not clear how far from the star TiO can expand before it is completely converted to TiO$_2$ or depleted into dust. 

The line profiles of TiO$_2$ are much more complex than those of TiO and display several emission components (Fig.\,\ref{Fig-aver}). The width of the combined profile of TiO$_2$ suggests that it reaches the terminal velocity. The emission is marginally resolved in our observations with the resolution of 0.9 arcsec, allowing the determination of its size of $r$\,$\approx$\,50\,R$_{\star}$.

Under the assumption of local-thermodynamic-equilibrium (LTE), we were able to constrain the rotational temperature of the TiO$_2$-bearing gas at $T_{\rm rot}$=250\,K. It was not possible to derive the excitation temperature of TiO. The line ratios suggest TiO is excited under non-LTE conditions. As mentioned, electronic bands of TiO have been observed in the optical spectrum of VY\,CMa, which provides evidence for radiative pumping and therefore LTE is violated. Interestingly, the submillimeter and optical emission lines of TiO have very similar profiles. For more details about the TiO and TiO$_2$ spectra see Kami\'nski \etal\ (\cite{kami_tio}).

\subsection{Aluminum monoxide in VY\,CMa}
\begin{figure*}\centering
\includegraphics[angle=270,width=\textwidth]{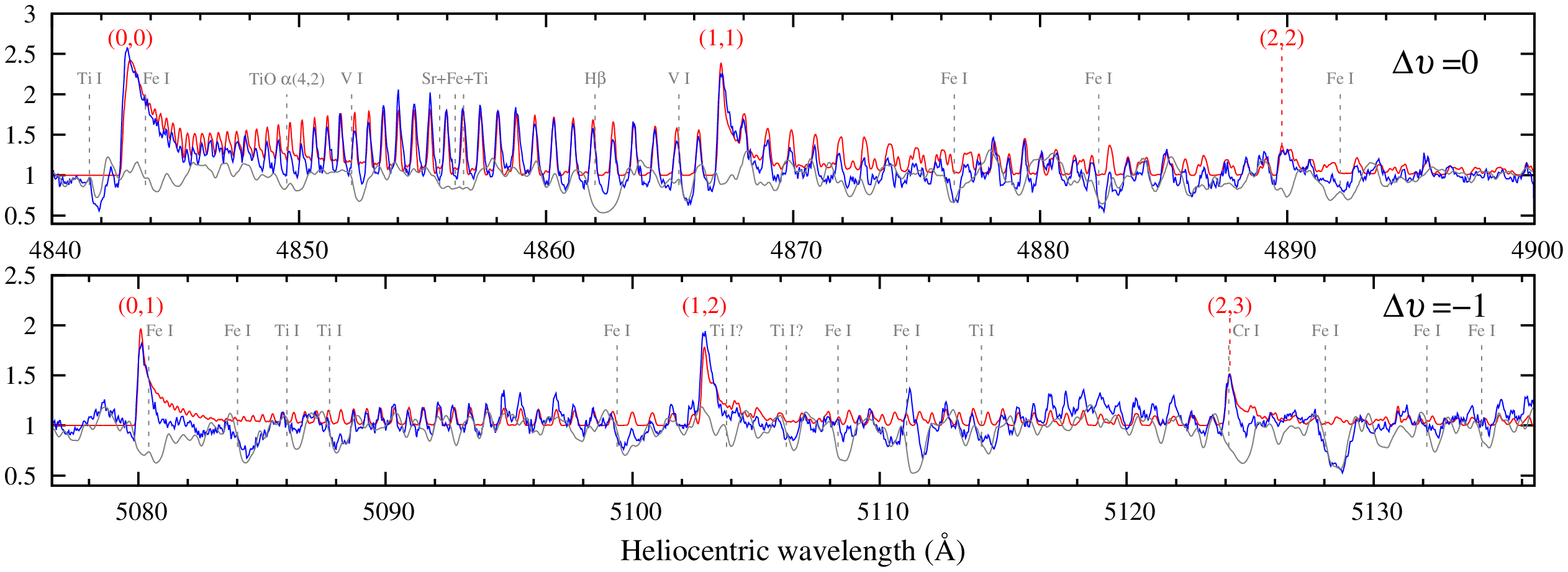}
\caption{The spectrum of VY\,CMa obtained with VLT/UVES (blue line) is compared to the simulation of the $B$--$X$ band of AlO in emission (red line); spectrum of Betelgeuse is also shown (gray line) to indicate the positions of photospheric and circumstellar absorption features. The ordinate is normalized flux density.}
\label{Fig-uves}
\end{figure*}

Encouraged by the discovery of the titanium oxides, we made an attempt to search for and characterize emission of AlO in VY\,CMa. Emission of this monoxide had been detected at millimeter wavelengths (Tenenbaum \& Ziurys \cite{TZ}) but at a very modest S/N ($\sim$3). AlO was also detected in our SMA survey (see below). However, at submillimeter wavelengths the rotational lines of AlO have hyperfine splitting that is of the order of the intrinsic line width, which complicates the analysis of the submillimeter profiles. Fortunately, the splitting is smaller at optical wavelengths motivating a search for AlO in this regime. We extracted an optical spectrum of VY\,CMa from the public ESO archive. It was obtained with the Ultraviolet and Visual Echelle Spectrograph (UVES) at the Very Large Telescope (VLT) in 2001. We found very strong emission features corresponding to the $B$--$X$ system of AlO at around 5000\,\AA\ (see Fig.\,\ref{Fig-uves}). This is the first identification of AlO electronic bands in this source, which is rather surprising considering that VY\,CMa has been a favorite source of many observers and was observed in the optical at many occasions.

We were able to model the optical emission of AlO including (to some degree) the effects of non-LTE excitation. Our simulations with rotational temperature of $T_{\rm rot}$\,=\,700\,K reproduce the observations very well, as can be seen in Fig.\,\ref{Fig-uves}. An analysis of profiles of individual rotational components allowed us to derive the intrinsic half-width of the emission of FWHM/2\,=\,7\,km\,s$^{-1}$, which implies that the AlO-bearing gas seen in the optical is located in the wind-acceleration zone, out to $r$\,$\approx$\,20\,R$_{\star}$. The results of the identification and the full analysis of the AlO bands in the optical spectrum of VY\,CMa are presented in Kami\'nski \etal\ (\cite{kami_alo}). 

After a careful analysis of the AlO lines detected in the SMA survey, whose profiles are  severely affected by hyperfine splitting, we found that with FWHM/2\,= 15\,km\,s$^{-1}$ the emission at submillimeter wavelengths is considerably broader than that seen in the optical. Additionally, the rotational diagram for AlO suggests that the lines observed with SMA are formed at an excitation temperature of $T_{\rm rot}$\,$\approx$\,400\,K which is lower than that found from the optical observations (700\,K). This suggests that the emission traced by the submillimeter emission arises from a more extended region around the star.    

\subsection{The oxides, condensation, and non-equilibrium chemistry}

Most of the attempts to describe the process of nucleation in circumstellar envelopes are based on the assumption of the chemical equilibrium. They predict (e.g., Gail \& Sedmayr \cite{GS}) that TiO, which is easily formed in photospheres of stars of our interest, is the dominant carrier of titanium in the outflow down to temperatures of about 1400\,K. In a reaction of TiO with water, TiO$_2$ is formed eventually consuming all TiO. At slightly lower temperatures more complex titanium oxides are formed (Ti$_2$O$_3$, Ti$_3$O$_5$, {\it etc.}) which grow and nucleate. Condensation should deplete the titanium-bearing species very effectively so they should be virtually absent at temperatures lower than 1000\,K (see Sharp \& Huebner \cite{SH}). Similarly, AlO is easily formed at temperatures above 2000\,K but is consumed in a reaction with water producing higher aluminum oxides. Owing to condensation, its abundance drops drastically with temperature starting from about 1800\,K (Sharp \& Huebner \cite{SH}). Therefore, from the observational point of view, at chemical equilibrium gas-phase AlO, TiO, and TiO$_2$ should be seen in thin layers above the photosphere and should be characterized by high ($>$1000\,K) excitation temperatures.

Our observational constraints on the excitation temperatures and the extent of the emission of the three oxides in VY\,CMa are inconsistent with the above picture -- the molecular emission is more extended and the gas is cooler; also the calculated molecular abundances, although uncertain by at least 1\,dex, are generally higher than those predicted in chemical equilibrium. This suggests that chemical equilibrium is violated in the circumstellar gas surrounding VY\,CMa and/or the condensation is less effective than assumed in the calculations.

It would be actually surprising if the gas around VY\,CMa was in chemical equilibrium. Studies of Humphreys \etal\ (\cite{H05,H07}) have shown that the complex structure of the nebula of VY\,CMa was formed by episodic mass-loss events (or ejections) which are violent enough to shock the material around the star. These episodes of enhanced mass loss are localized (i.e., non-spherical) and likely related to the convective activity of this extreme supergiant. Whether the same type of activity is taking place now and whether it affects the gas traced by the emission of the refractory oxides (i.e., at $r\,\leq$\,50\,R$_{\star}$) should be verified by observations at very high angular resolution. %Observations of other molecules thought to be formed in the inner outflow also indicate presence of non-equilibrium chemistry.    

To understand the chemistry and dust condensation in evolved stars, and in supergiants in particular, non-equilibrium models are desperately needed. Cherchneff (e.g., \cite{cherchneff}) and others have started to calculate such more realistic models, but these studies are limited to stars on the asymptotic giant branch with a special focus on carbon stars. We hope that the new observations of VY\,CMa presented here will encourage theoreticians to perform chemical studies adjusted to physical conditions around red supergiants.

The detection of the three oxides in VY\,CMa allows for an observational examination of the dust nucleation scenarios for the first time. With its violent circumstellar environment, however, VY\,CMa may be an exceptional case, not representative for the broad range of dust-forming stars, and therefore the search for the three species should continue in other objects. 

%------------------------------------------------------------
\section{Can we detect the refractory oxides in Betelgeuse?}
\begin{figure*}\centering
\includegraphics[angle=270,width=\textwidth]{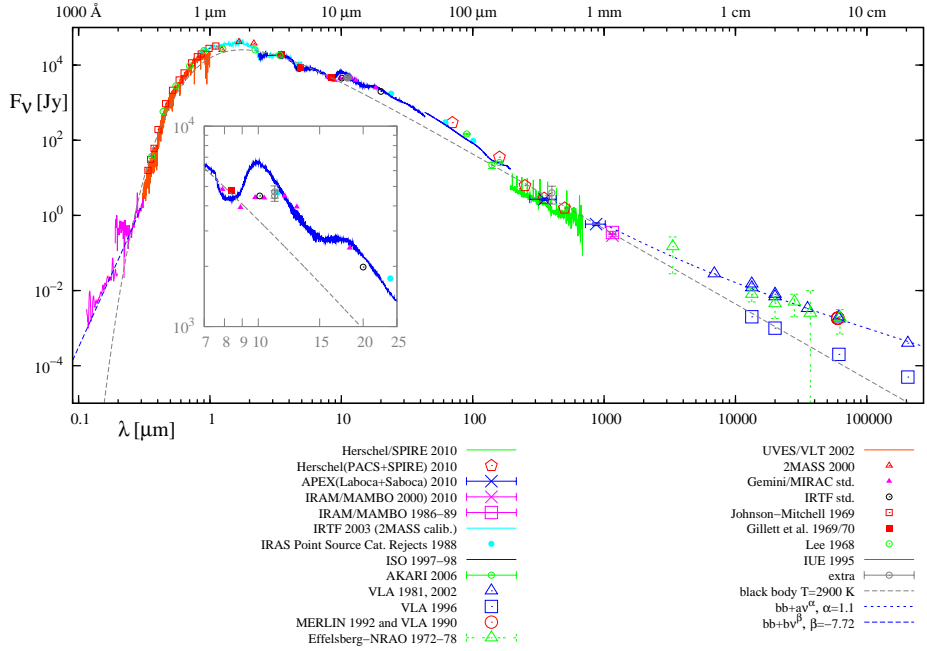}
\caption{The spectral energy distribution of Betelgeuse in a broad spectral range compiled from literature data (Kami\'nski \etal, in preparation). An infrared excess owing to thermal dust emission is seen between 8 and 200\,$\mu$m.}
\label{Fig-sed}
\end{figure*}

With the copious amounts of material and the chemical richness, the nebula of VY\,CMa is probably an extreme example of circumstellar environment of a red supergiant.  Betelgeuse ($\alpha$\,Orionis) constitutes its counter-example. With a mass-loss rate of only $\sim$10$^{-6}$\,M$_\odot$\,yr$^{-1}$ and low gas-to-dust mass ratio of 200--1700 (Harper \etal\ \cite{harper}), compared to VY\,CMa Betelgeuse appears to be a ``naked" red supergiant. Nevertheless, it is known to possess a MOLsphere above its photosphere and to exhibit weak molecular emission of CO, SiO, OH, and H$_2$O at centimeter to submillimeter wavelengths, so it has some molecular gas in the circumstellar environment. The presence of dust is manifested best by an infrared excess between 8 and 200\,$\mu$m (Fig.\,\ref{Fig-sed}). Observations of the dust-forming molecules would help to understand why the molecular and dusty envelope of Betelgeuse is so much poorer than that of VY\,CMa. Here, we briefly discuss the possibilities to observe the molecules important for the nucleation process, mainly TiO and AlO, in the wind of Betelgeuse.  

Betelgeuse has been observed in the optical with high-resolution spectrographs on many occasions, yet it has never been reported as a source of optical molecular emission. We examined a spectrum of Betelgeuse obtained with UVES (Bagnulo \etal\ \cite{pop}) at S/N of up to $\sim$300 and found no signs of molecular emission. Betelgeuse does exhibit strong absorption bands of TiO, as expected for a star of spectral type M\,1--2, and does not show any sign of photospheric AlO, which is also consistent with its spectral type (AlO is seen in spectral types later than M4). In fact, we used the spectrum of Betelgeuse as a reference AlO-free spectrum in our analysis of the AlO bands in VY\,CMa (see Fig.\,\ref{Fig-uves}). 

Optical emission of TiO and AlO is seen only in sources where the central continuum source is heavily obscured by circumstellar dust (e.g., Kami{\'n}ski \etal\ \cite{kami_v4332}). Although the AlO bands in VY\,CMa have an intensity 2.5 times higher than that of the observed local continuum, their true intensity relative to the photospheric flux is unknown. Owing to extinction in the extended nebula ($A_V$\,$\approx$\,3.2\,mag), the observed continuum represents only a small fraction ($\sim$5\%) of the original flux leaving the photosphere. The flux ratio of the molecular emission and the photosphere is  close to the observed value only if the AlO emission arises close to the photosphere and is affected by the same extinction. Let us assume this is the case. With the mass-loss rate of Betelgeuse of $\sim$10$^{-6}$\,M$_\odot$\,yr$^{-1}$ and the same AlO abundance as in VY\,CMa, the intensity of the molecular emission relative to the photospheric flux would be two orders of magnitude lower than in VY\,CMa, giving $\sim$2.5\% of the continuum flux. If the circumstellar abundances of AlO (with respect to hydrogen) are lower in Betelgeuse than in VY\,CMa, this value would be even smaller. Thus, the potential molecular emission would be detectable in spectra with S/N higher than $\sim$40. Spectra of much better quality exist showing no signs of molecular emission, suggesting that either AlO has a very low abundance (at least a few times lower than in VY\,CMa) or does not exist at all in the wind of Betelgeuse. 

The above conclusion, however, does not seem to be consistent with the results of Verhoelst \etal\ (\cite{al2o3}). They claimed that amorphous alumina (Al$_2$O$_3$) exists in the outflow and extends from the stellar surface up to a few stellar radii. This finding was based on simulations of the spectrum of Betelgeuse around 9\,$\mu$m obtained with the Infrared Space Observatory. Since Al$_2$O$_3$ is formed  from AlO, which does not exist in the photosphere, AlO should be present somewhere in the inner outflow. That it is not seen in the UVES spectrum is therefore puzzling. The proposal of Verhoelst \etal\ (\cite{al2o3}) should be verified by independent studies of the infrared spectra. 
 
The Atacama Large Millimeter Array (ALMA), with its unprecedented sensitivity, opens another possibility for detecting the species important for dust formation at (sub-)millimeter wavelengths. Even with ALMA, though, long integrations would be necessary to conclusively test for the presence of dust-forming species in Betelgeuse.

\end{document}